\documentclass[twocolumn,showpacs,preprintnumbers,amsmath,amssymb,prl]{revtex4}

\usepackage[dvips]{graphicx}
\usepackage{dcolumn}
\usepackage{amssymb}
\usepackage{amsfonts}
\usepackage{bm}

\begin{document}

\preprint{APS}
\title{Quasi-localization and quasi-mobility edge for  light  atoms mixed with  heavy ones}
\author{E. Kogan}
\email{kogan@mail.biu.ac.il}
\affiliation{Jack and Pearl Resnick
Institute, Physics Department, Bar Ilan University, Ramat Gan 52900,
Israel}
\date{\today}

\begin{abstract}
A mixture of light and heavy  atoms is considered. We study the kinetics
of the light atoms,  scattered by the heavy ones, the latter undergoing slow
diffusive motion. In  three-dimensional space we claim the
existence of a crossover region (in energy), which
separates the states of the light atoms with  fast diffusion and the
states with slow diffusion; the latter is determined by the
dephasing time. For the two dimensional case we have a transition
between  weak localization, observed when the dephasing length is
less than the localization length (calculated for static
scatterers), and strong localization observed in the opposite
case.
\end{abstract}
\maketitle

Mixtures of different species of cold atoms present an interesting
field of many particle physics.  Two or more different types of
atoms can be mixed, where one type of atoms can be relatively light
(e.g. $^6$Li), and the other type is heavy (e.g. $^{87}$Rb). Quantum
tunneling of light atoms is a phenomenon, interesting both from an
experimental and theoretical point of view. The heavy atoms serve as
slow moving scatterers for the light atoms.   Lately, it was realized that ultracold atomic
gases appear very convenient for experimental studies of Anderson
localization of the light atoms, both for the case of Bose-Einstein condensates, and
for fermionic gases
\cite{gavish,paredes,clement,fort,schulte,kuhn,ziegler,ospelkaus,clement2,massignan,sanches}.

Kinetics of classical particles in a disordered medium can be
described by the Boltzmann equation. The most drastic manifestation of the difference
between the kinetics of classical particles and that of quantum
ones is Anderson localization. It is well known that for $d=1$ and
$d=2$, where $d$ is the dimensionality of space, all the states are
localized, and for $d=3$ there  exists a mobility edge $E_c$, 
the energy which separates the states with finite diffusion
coefficient and states with the diffusion coefficient being exactly
equal to zero. (For reviews, see e.g. Refs.
\cite{kramer,chakravarty,rammer}.) All this is true provided the
disorder is static. A natural question arises: what happens with this
picture when the scatterers slowly move?

To answer this question  we need some quantitative theory of localization. As
such we will use the self-consistent localization theory by Vollhard
and W\"olfle \cite{vollhardt}. Of crucial importance in the above
mentioned theory are maximally crossed diagrams (the sum of all such
diagrams is called the Cooperon) for the two-particle Green function.
The calculations of these diagrams for the case of moving scatterers
were done in the paper by Golubentsev \cite{golubentsev}.

One should notice that  we consider the heavy atoms as classical
objects whose diffusive motion is not affected by localization
effects. On the other hand, we consider the light atoms as quantum
objects. Thus the temperature of the atom gases should satisfy the
inequalities \cite{landau}
\begin{eqnarray}
\frac{\hbar^2}{M}N^{2/d}\ll T \ll \frac{\hbar^2}{m}n^{2/d},
\end{eqnarray}
where $M$ and $N$ are the mass and concentration of heavy atoms
respectively, $m$ and $n$ are the mass and concentration of light
atoms and $T$ is the temperature. The large ratio between the masses
of the two types of atoms considered is crucial  for the
applicability of the methods used in this work also because
following Ref. \cite{golubentsev}, we shall ignore the change of the
energy of the light atoms as a result of a scattering by a heavy
one.

In the first part of the present paper we reproduce the
results by Golubentsev (trivially generalized for the arbitrary
dimensionality of space). In the second part we use the results for
the Cooperon as an input for the self-consistent localization
theory, which we  modify to take into account the slow motion of
scatterers. In the third part we discus the results obtained.

The quantum particles are scattered by the potential
\begin{eqnarray}
V(r,t)=V\sum_a\delta\left(r-r_a(t)\right).
\end{eqnarray}
Define the correlator
\begin{eqnarray}
K(r-r',t-t')=<V(r,t)V(r',t')>.
\end{eqnarray}
To leading order in the density of scatterers  we have for
the Fourier component of the correlator
\begin{eqnarray}
K(q,t)=V^2\left\langle\int\exp\left\{ (i{\bf q}({\bf r}-{\bf r}') \right\}\right.\nonumber\\
\left. \times  drdr'\sum_a\delta\left({\bf r}-{\bf r}_a(t)\right)
\sum_{a'}\delta\left({\bf r}'-{\bf r}_{a'}'(0)\right)\right\rangle\nonumber\\
=V^2\sum_a\left\langle\exp\left\{ iq({\bf r}_a(t)-{\bf
r}_a(0))\right\}\right\rangle =nV^2f({\bf q},t),
\end{eqnarray}
where $n$ is the scatterer density. We consider the case when the
scatterers undergo slow diffusive motion. In the ballistic case
\begin{eqnarray}
f({\bf q},t)=\exp\left(-\frac{{\bf q}^2T}{2M}t^2\right),\qquad
|t|\ll\tau_{imp},
\end{eqnarray}
In the diffusive case
\begin{eqnarray}
f({\bf q},t)=\exp\left(-\frac{{\bf
q}^2T\tau_{imp}}{2M}|t|\right),\qquad |t|\gg\tau_{imp},
\end{eqnarray}
where we have used the fact that
\begin{eqnarray}
<{\bf v}_{imp}^2>=\frac{dT}{M},
\end{eqnarray}
and $\tau_{imp}$ is the  mean free time of the scatterers.

For the Cooperon  we get \cite{golubentsev}
\begin{eqnarray}
\label{cooperon}
C_E({\bf q})= \int_0^{\infty}\exp\left\{-D(E)q^2t
-\frac{1}{\tau} \int_0^t(1-f_{t'})dt'\right\}dt,
\end{eqnarray}
where $E$ is the energy of each of  the two quantum particle lines in the
Cooperon diagram, and $q$ is the sum of their momenta (see Fig. 1).
Also
\begin{eqnarray}
\frac{1}{\tau}=\left\{\begin{array}{ll}nV^2k^2/\pi v & d=3\\
                                       nV^2k/v & d=2\\
                                       nV^2/v & d=1\end{array}\right..
\end{eqnarray}
We'll assume that $\tau\ll \tau_{imp}$.  The quantity $f_t$ is
$f({\bf k})$ averaged with respect to the iso-energetic surface. We
obtain
\begin{eqnarray}
f_t=\left\{\begin{array}{ll}y_d\left(\frac{t^2}{\tau_{\lambda}^2}\right) & |t|\ll\tau_{imp} \\
                         y_d\left(\frac{|t|\tau_{imp}}{\tau_{\lambda}^2}\right) & |t|\gg\tau_{imp}\end{array}\right.
\end{eqnarray}
where
\begin{eqnarray}
\tau_{\lambda}=\left(\frac{2k^2T}{M}\right)^{-1/2}.
\end{eqnarray}
For $d=3$,
$y_3(x)=(1-e^{-x})/x$ \cite{golubentsev}.
For $d=2$
\begin{eqnarray}
f_t=\int\frac{d{\bf s}'}{2\pi}f(k(s-s'),t).
\end{eqnarray}
Using the integral
\begin{eqnarray}
\frac{1}{\pi}\int_0^{\pi}d\theta e^{-A(1-\cos\theta)}=e^{-A}I_0(A),
\end{eqnarray}
where $I_0$ is the modified Bessel function, we obtain
\begin{eqnarray}
y_2(x)=e^{-x/2}I_0(x/2).
\end{eqnarray}
For $d=1$
\begin{eqnarray}
y_1(x)=e^{-x/2}.
\end{eqnarray}

Eq. (\ref{cooperon}) can be easily understood if we compare diagrams
for the Diffuson (the sum of all ladder diagrams) and the Cooperon
in Fig. 1. The Diffuson does not have any mass because of the Ward
identity. In the case of the Cooperon, the Ward identity is broken,
and the difference $[1-f(t)]$ shows how strongly. The interaction
line which dresses the single particle propagator is given by the static
correlator, and the interaction line which connects two different
propagators in a ladder is given by the dynamic correlator.
\begin{figure}
\includegraphics[angle=0,width=0.45\textwidth]{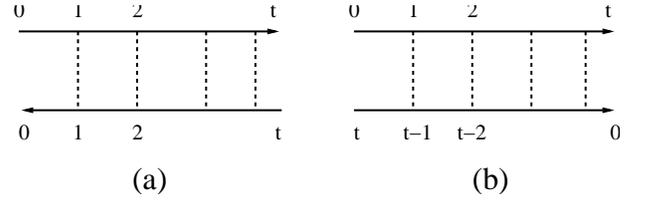}
\caption{Diagrams for the Diffuson (a) and the Cooperon (b). Solid
line is dressed quantum particle propagator, dashed line connecting points
${\bf r},t$ and ${\bf r}',t'$ corresponds to $K({\bf r-r'},t-t')$.}
\end{figure}
The time-reversal invariance in the system we are considering is
broken due to dephasing; the diffusion pole of the particle-particle
propagator disappears, although the particle-hole propagator still has a
diffusion pole, which is guaranteed by particle number conservation.
(The weak localization effects for the case of inelastic
electron-phonon  scattering  were discussed by Afonin et al.
\cite{afonin}.)

Considering the limiting cases, from Eq. (\ref{cooperon}) we obtain:

\noindent
(i) in the  case $2k^2T/M\ll \tau/\tau_{imp}^3$
\begin{eqnarray}
\label{cooperon22}
 C_E({\bf q})=
\int_0^{\infty}\exp\left[-D(E){\bf
q}^2t-t^2/\tau_{\varphi}^2(E)\right]dt,
\end{eqnarray}
where
\begin{equation}
\label{g2}
\tau_{\varphi}=\left(\frac{2M}{k^2T}\frac{\tau}{\tau_{imp}}\right)^{1/2};
\end{equation}
(ii) in the  case $\tau/\tau_{imp}^3\ll 2k^2T/M\ll 1/\tau^2$
\begin{eqnarray}
\label{cooperon2} C_E({\bf q})= \int_0^{\infty}\exp\left[-D(E){\bf
q}^2t-t^3/\tau_{\varphi}^3(E)\right]dt,
\end{eqnarray}
where
\begin{equation}
\label{g}
\tau_{\varphi}=\left(\frac{3M\tau}{k^2T}\right)^{1/3}.
\end{equation}
Thus we obtain the crucial parameter  - the dephasing time
$\tau_{\varphi}$.

The results  for the dephasing time (up to a numerical factors of
order  one) can be understood using simple qualitative arguments.
Consider the ballistic regime. If a single collision leads to the
quantum particle energy change $\delta E$, the dephasing time could
be obtained using Eq. \cite{altshuler}
\begin{equation}
\label{toy0}
\tau_{\varphi}\delta
E\sqrt{\frac{\tau_{\varphi}}{\tau}}\sim 2\pi,
\end{equation} where $\tau_{\varphi}/\tau$ is just the
number of scatterings during the time  ${\tau_{\varphi}}$. So in
this case
\begin{equation}
\label{toy}
\frac{1}{\tau_{\varphi}^3}\sim\frac{(\delta E)^2}{\tau}.
\end{equation}
If we notice that $1/\tau_{\lambda}$ is the averaged quantum
particle energy change in a single scattering, $\delta E$, we
immediately regain Eq. (\ref{g}). Eqs. (\ref{toy0}) and (\ref{toy})
also imply that if the scattering is quasi elastic (and slow
motion of scatterers means just that), the energy relaxation time is
much larger than the dephasing time \cite{altshuler}. Hence we have
the right to ignore the Doppler caused cumulative energy shift, which
otherwise would have lead to the appearance of the Diffuson mass.

Inserting Eq. (\ref{cooperon2}) into the self-consistent equation,
for the diffusion coefficient $D$  we obtain 
\begin{eqnarray}
\label{ret}
 \frac{D_0(E)}{D(E)}=1+\frac{1}{4\pi^2 mk} \sum_{\bf
q}C_E({\bf q})
\end{eqnarray}
where $D_0$ is the diffusion coefficient calculated in the Born
approximation
\begin{eqnarray}
\label{diffusion}
 D_0=\frac{1}{d}v^2\tau;
\end{eqnarray}
$v$ is the particles velocity, and the momentum cut-off $|{\bf
q}|<1/\ell$ is implied, where $l=k\tau/m$ is the mean free path.
 Thus we obtain
\begin{eqnarray}
\label{re}
\frac{D_0}{D}=1+\frac{1}{\pi mk}
 \int_0^{\infty}dt
\int_0^{1/l}dq\; q^{d-1}\nonumber\\
\times\exp\left[-Dq^2t -g(t/\tau_{\varphi}) \right],
\end{eqnarray}
where $g(x)$ is some function which goes to infinity when $x$ goes
to infinity as some power of $x$ higher than one (in the
particular case of ballistic regime $g(x)=x^3$, and in the diffusive
regime $g(x)=x^2$.)

Introducing dimensionless variables we obtain
\begin{eqnarray}
\label{re4}
\frac{D_0}{D}=1+\frac{1}{\pi}\frac{1}{(kl)^{d-1}}
\int_0^{\infty}d\tilde{t} \int_0^{1}d\tilde{q}\; \tilde{q}^{d-1}\nonumber\\
\times\exp\left[-\frac{1}{d}\frac{D}{D_0}\tilde{q}^2\tilde{t}
-g(\tilde{t}\tau/\tau_{\varphi}) \right].
\end{eqnarray}
Thus we have obtained an algebraic equation for $D/D_0$, which (equation)
depends upon two parameters: $\tau_{\varphi}/\tau\gg 1$ and $kl$,
which can be arbitrary.

Let us start analysis of this equation with the case $d=2$.
Calculating the integral with respect to $\tilde{q}$ we obtain
\begin{eqnarray}
\label{re5}
\frac{D}{D_0}=1-\frac{1}{\pi kl} \int_0^{\infty}
\frac{d\tilde{t}}{\tilde{t}}\left[1-
e^{-\frac{D\tilde{t}}{2D_0}}\right]
e^{-g(\tilde{t}\tau/\tau_{\varphi})}.
\end{eqnarray}
Let us make the assumption (which we'll justify a posteriori)
\begin{equation}
\label{assumption}
D\tau_{\varphi}/D_0\tau\gg 1.
\end{equation}
 To calculate the integral
\begin{eqnarray}
I(\lambda)=\int_0^{\infty}
\frac{d\tilde{t}}{\tilde{t}}\left[1-e^{-\lambda \tilde{t}}\right]
e^{-g(\tilde{t})},\qquad \lambda\gg 1,
\end{eqnarray}
let us divide the region of integration by choosing some $x$ satisfying $1/\lambda\ll x\ll 1$. We obtain
\begin{eqnarray}
\label{re55}
I(\lambda)=\left[\int_0^{x}+\int_x^{\infty}\right]
\frac{d\tilde{t}}{\tilde{t}}\left[1-e^{-\lambda \tilde{t}}\right]
e^{-g(\tilde{t})}\nonumber\\
=\int_0^{x}
\frac{d\tilde{t}}{\tilde{t}}\left[1-e^{-\lambda\tilde{t}}\right]
+\int_x^{\infty}
\frac{d\tilde{t}}{\tilde{t}}e^{-g(\tilde{t})}\nonumber\\
=\ln (\lambda x)-\ln x=\ln\lambda.
\end{eqnarray}
(In Eq. (\ref{re55}) we have ignored all numerical factors of order 1
in the argument of the
logarithms.) Hence, Eq. (\ref{re5}) can be presented in the form
\begin{eqnarray}
\label{re6}
\frac{D}{D_0}=1-\frac{1}{\pi kl}
\ln\left(\frac{D\tau_{\varphi}}{D_0\tau}\right).
\end{eqnarray}
Solution of Eq. (\ref{re6}) is particularly simple in two limiting
cases: $l_{\varphi}\ll \xi$ and $l_{\varphi}\gg \xi$, where
$l_{\varphi}=v\tau_{\varphi}$ is the  dephasing length, and
$\xi=le^{\frac{\pi kl}{2}}$ is the localization length
\cite{rammer}. In the former case we obtain just weak localization
corrections
\begin{eqnarray}
\label{re7}
\frac{D}{D_0}=1-\frac{1}{\pi kl}
\ln\frac{\tau_{\varphi}}{\tau},
\end{eqnarray}
and in the latter case
\begin{eqnarray}
\label{2d}
D=\frac{\xi^2}{\tau_{\varphi}}.
\end{eqnarray}
We see that in both cases the assumption (\ref{assumption}) is
satisfied.

Results of a numerical solution of Eq. (\ref{re5}) for $g(x)=x^3$, $g(x)=x^2$ and $g(x)=x$ are presented on Fig. 2.
One can see that the curves for $D/D_)$ are practically indistinguishable. Thus the exact form of the function $g(x)$ is not important. All the relevant information is
contained in the  dephasing time, determined by the
parameter $\tau_{\varphi}$.

\begin{figure}
\includegraphics[angle=0,width=0.45\textwidth]{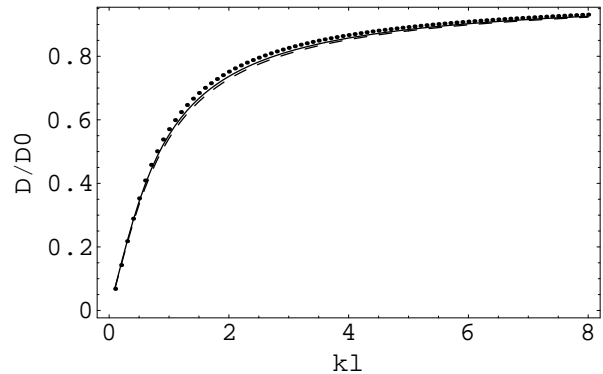}
\caption{The results of numerical solution of Eq. (\ref{re5})
($\tau_{\varphi}/\tau=10$) for $g(x)=x^2$ (solid line), $g(x)=x^3$ (dashed line), and $g(x)=x$ (dots).}
\end{figure}

 Notice, that the quantum diffusion of particles scattered by
the slow moving scatterers turns out to be similar to the case when
there are two separate scattering mechanisms: strong elastic
scattering  causing relaxation of momentum, and weak  inelastic
scattering due to say, phonons, causing dephasing (except for the
definition of $\tau_{\varphi}$). Strong dependence of the diffusion
coefficient for $d=2$ upon the ratio of the dephasing and the
localization length (for the case of two scattering mechanisms) was
thoroughly discussed in Refs. \cite{gogolin,gogolin2,minkov,kuhn}.

As it was noticed by Gogolin and Zimanyi \cite{gogolin}, there is a
lower bound of temperature for the validity of Eq. (\ref{2d}.  At  low
enough  temperatures variable range hopping, which is of course not taken into
account by the self-consistent localization theory is the main
diffusion mechanism. So Eq. (\ref{re}) is valid, provided
\begin{eqnarray}
\label{vrh}
T\gg\Delta E,
\end{eqnarray}
where $\Delta E$ is the average energy difference between
neighboring localized states. Eq. (\ref{vrh}) can be presented as
\begin{eqnarray}
\label{vrh2}
T\gg \frac{1}{ml^2}e^{-\pi kl}.
\end{eqnarray}
On the other hand, inequality $l_{\varphi}\gg \xi$ after
substitution of Eq. (\ref{g2}) gives
\begin{eqnarray}
\label{vrh3}
 T\ll\frac{M}{m}\frac{\tau}{\tau_{imp}} \frac{1}{ml^2}e^{-\pi kl},
\end{eqnarray}
and after substitution of Eq. (\ref{g}) gives
\begin{eqnarray}
\label{vrh4}
 T\ll\frac{M}{m} \frac{1}{ml^2}e^{-\pi kl}.
\end{eqnarray}
We again obtain see the importance of the large parameter $M/m$.

In fact, Eq. (\ref{2d}) is valid both for $d=1$ and $d=3$ (in the
latter case, provided we have localization in the absence of
dephasing). Taking into account the numerical results obtained for $d=2$, for the purpose of
semi-quantitative  analysis we may approximate Eq. (\ref{re}) by
\begin{eqnarray}
\label{reap}
\frac{D_0}{D}=1+\frac{1}{\pi mk}
 \int_0^{\infty}dt
\int_0^{1/l}dq\; q^{d-1}\nonumber\\
\times\exp\left[-Dq^2t -t/\tau_{\varphi} \right].
\end{eqnarray}
Calculating the integral with respect to $t$ we obtain Eq. (\ref{reap}) in the form
\begin{eqnarray}
\frac{D}{D_0}=1-\frac{d}{\pi(kl)^{d-1}}\int_0^1\frac{d\tilde{q}\tilde{q}^{d-1}}{\tilde{q}^2+\frac{l^2}{D\tau_{\varphi}}}.
\end{eqnarray}

For $d=2$ we obtain 
\begin{eqnarray}
\frac{D}{D_0}=1-\frac{1}{\pi kl}\ln\left[\frac{D\tau_{\varphi}}{l^2}+1\right],
\end{eqnarray}
which in our approximation coincides with Eq. (\ref{re6}).

For $d=1$ we obtain from Eq. (\ref{reap})
\begin{eqnarray}
\frac{D}{D_0}=1-\frac{1}{\pi}\frac{\sqrt{D\tau_{\varphi}}}{l}\tan^{-1}\frac{\sqrt{D\tau_{\varphi}}}{l}.
\end{eqnarray}
Again ignoring  numerical multipliers of order 1 we obtain
\begin{eqnarray}
\label{1d}
D= D_0\frac{\tau}{\tau_{\varphi}}
\end{eqnarray}
If we take into account that for $d=1$ we have $\xi\sim l$, we see that Eq. (\ref{1d}) is equivalent to Eq. (\ref{2d}).
One must admit, however, that for $d=1$ the self-consistent
localization theory should be handled with care. In addition
interaction between quantum particles, not considered in the present
paper, may strongly influence the localization processes
\cite{gornyi}.

For $d=3$ from Eq. (\ref{reap}) we obtain
\begin{eqnarray}
\label{3d}
\frac{D}{D_0}=1-\frac{3}{\pi(kl)^2}\left[1-\frac{l}{\sqrt{D\tau_{\varphi}}}\tan^{-1}\frac{\sqrt{D\tau_{\varphi}}}{l}\right].
\end{eqnarray}
One can see, that for $d=3$ (similar to the case $d=2$) Eq. (\ref{2d}) ceases to be valid
when the localization length $\xi$ becomes large enough, which happens when the parameter $kl$
approaches the critical value $\sqrt{3/\pi}$ from below. In fact, in this region Eq. (\ref{3d}) can be presented as
\begin{eqnarray}
\label{3da}
\frac{D}{D_0}=2\sqrt{3\pi}(\lambda_c-\lambda)+\frac{l}{\sqrt{D\tau_{\varphi}}}
\tan^{-1}\frac{\sqrt{D\tau_{\varphi}}}{l},
\end{eqnarray}
where $\lambda=1/\pi kl$. and $\lambda_c=1/\sqrt{3\pi}$.
After assuming that the term $2\sqrt{3\pi}(\lambda_c-\lambda)$  can be ignored with respect to the second term
in the rhs of Eq. (\ref{3da}), and that $D\tau_{\varphi}/l\gg1$
we obtain
\begin{eqnarray}
\label{third}
D=\frac{l^2}{\tau^{2/3}\tau_{\varphi}^{1/3}}.
\end{eqnarray}
Now checking  the  assumptions and taking into account that in the critical region \cite{rammer}
$\xi=l/|\lambda-\lambda_c|$, we see that Eq. (\ref{third}) is valid, provided
\begin{eqnarray}
\xi>l^{2/3}l^{1/3}_{\varphi}.
\end{eqnarray}
The results of numerical solution of Eq. (\ref{3d}) are presented on Fig. 3.
\begin{figure}
\includegraphics[angle=0,width=0.45\textwidth]{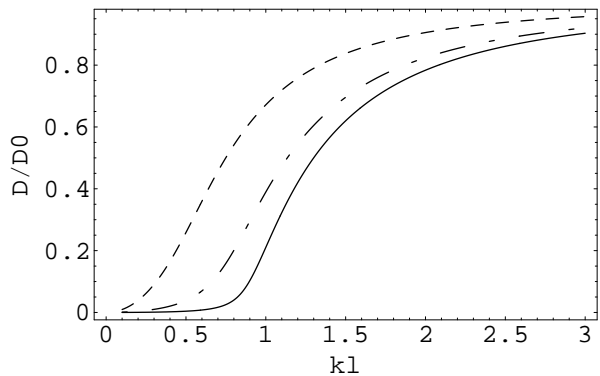}
\caption{The results of numerical solution of Eq. (\ref{3d}) for
$\tau_{\varphi}/\tau=10$ (dashed line), $\tau_{\varphi}/\tau=100$ (dot-dashed line), and  $\tau_{\varphi}/\tau=1000$ (solid line).}
\end{figure}

Notice that in accordance with Refs. \cite{gogolin,gogolin2} the
dephasing time dependence of the diffusion coefficient can be
obtained from its frequency dependence by replacing $\omega$ by
$i\tau_{\varphi}$. Eq. (\ref{re}) in the absence of dephasing but
for finite frequency is \cite{rammer}
\begin{eqnarray}
\label{re10} \frac{D_0}{D(\omega)}=1+\frac{1}{\pi mk}
 \int_0^{\infty}dt
\int_0^{1/l}dq\; q^{d-1}\nonumber\\
\times\exp\left[-D(\omega)q^2t +i\omega t) \right].
\end{eqnarray}
The localization length is defined \cite{rammer} as
\begin{equation}
\xi=\lim_{\omega\to 0}\sqrt{\frac{D(\omega)}{-i\omega}}.
\end{equation}
Analyzing the solution qualitatively, we may substitute
$1/\tau_{\varphi}$ for $-i\omega$ into the definition of the
localization length (\ref{re10}) and obtain Eq. (\ref{2d}).

\section{Conclusions}

We considered the influence of slow random motion of random
scatterers on the localization of quantum particles. It turned out
that whenever the states of the quantum particles were localized,
under the assumption, that the same scatterers are  {\it static},
taking the motion of the scatterers into account leads to a finite
value of the diffusion coefficient.  In particular, for the three
dimensional  case, there exists a narrow crossover region in energy
space, which separates the states with high and low diffusion
coefficient, the latter being inversely proportional to the
dephasing time. (For the states with fast  diffusion  the dephasing
is irrelevant.) Like the position of the mobility edge in the case
of static scatterers, the position of this crossover region is
determined by the criterion that the mean free path  is of the
order of the quantum particle wavelength. This crossover region we
call the quasi-mobility edge, and the phenomena in general we call
quasi-localization.  For the two dimensional case we have a
transition between  weak localization, observed when the
dephasing length is less than the localization length (calculated
for static scatterers), and strong localization observed in the
opposite  case.

The main application of our results  we see as lying in the
description of kinetics of ultracold gases.
However, we would like to mention  possible application of these
results to at least one other field. In our previous publication
\cite{kog}, we studied the influence of dephasing on the Anderson
localization of the electrons in magnetic semiconductors, driven by
spin fluctuations of magnetic ions. There the role of heavy
particles was played by magnons; complete spin polarization of
conduction electrons prevented magnon emission or absorption
processes, and only the processes of electron-magnon scattering
being allowed.

\end{document}